# Characteristics of Respiratory Microdroplet Nuclei on Common Substrates

Alexandros Kosmidis-Papadimitriou,[1] Shaojun Qi,[1] Ophelie Squillace,[1] Nicole Rosik,[1] Mark Bale,[2] Peter J. Fryer,[1] and Zhenyu J. Zhang[1,*]

[1]School of Chemical Engineering, University of Birmingham, Edgbaston, Birmingham, B15 2TT United Kingdom
[2]DoDxAct Ltd, Wells, BA5 2LW, United Kingdom

## Abstract

To evaluate the role of common substrates in the transmission of respiratory viruses, in particularly SARS-CoV-2, uniformly distributed microdroplets (~10 μm diameter) of artificial saliva were generated using an advanced inkjet printing technology to replicate the aerosol droplets, and subsequently deposited on five substrates, including glass, PTFE, stainless steel, ABS, and melamine. The droplets were found to evaporate within a short timeframe (less than 3 seconds), which is consistent with previous reports concerning the drying kinetics of picolitre droplets. Using fluorescence microscopy and atomic force microscopy, we found that the surface deposited microdroplet nuclei present two distinctive morphological features as the result of their drying mode, which is controlled by both interfacial energy and surface roughness. Nanomechanical measurements confirm that the nuclei deposited on all substrates possess similar surface adhesion (~ 20 nN) and Young's modulus (~4 MPa), supporting the proposed core-shell structure of the nuclei. We suggest that appropriate antiviral surface strategies, e.g. functionalisation, chemical deposition, could be developed to modulate the evaporation process of microdroplet nuclei, and subsequently mitigate the possible surface viability and transmissibility of respiratory virus.

## Introduction

Transmission of respiratory viruses can take place in different modes, either directly via contact between individuals, indirectly via commonly touched objects or surfaces, or directly through the air in the form of large droplets or small aerosols [1]. Surface transmission, in particularly via surface fomite, was viewed as one of the primary concerns since the initial stages of the pandemic caused by Severe Acute Respiratory Syndrome coronavirus-2 (SARS-CoV-2) in 2020 [2]. One of the early studies suggested that SARS-CoV-2 remain viable in aerosols for at least three hours, and that SARS-CoV-2 is more stable on plastic and stainless steel than on copper and cardboard [3], which highlighted the unique role of the common surfaces in virus transmission during the pandemic. The impact of substrate on the surface viability of virus was further demonstrated with a large range of Middle East Syndrome coronavirus (MERS-CoV) [4].

There has been compelling arguments that the transmission of SARS-CoV-2 after touching surfaces should be considered as relatively minimal [5], given that it is improbable that an infected person coughs or sneezes on a surface (with sufficient quantity of infectious virus), and someone else touches that surface shortly after (within 1-2 hour) [6]. This rationale is sound and sensible, on the assumption that surface transmission takes place via a large quantity of respiratory fluid and that the virus would be inactivated beyond the timeframe suggested. An extensive list of evidence, including superspreading events, long-range transmission, asymptomatic transmission, was given to support that the dominating transmission route of SARS-CoV-2 is airborne [7]. Nevertheless, the possible presence of infectious virus on solid substrates, in particular on high-touch environmental surfaces, could have significant implications for both social and healthcare practice.

There remain significant knowledge gaps in drawing conclusions on the possible role of surface in preserving and transmitting SARS-CoV-2 viruses due to the multitude complexities involved. This is highlighted in a systematic review by Onakpoya and colleagues focusing on the role of fomite transmission over 64 studies [8], which concluded that no evidence is available to confirm viral infectivity or transmissibility via fomites, but that none of the studies surveyed is sufficiently methodologically robust to adequately address the question. A similar reflection questioning the unlikelihood of indirect transmission through contaminated surfaces has been reported recently [9]. A core element that underpins the inconsistent viewpoints on the transmission pathways of SARS-CoV-2 is the physico-chemical properties of the microdroplets. Other than the discrepancy over the definition of aerosols, droplets, particles, and droplet nuclei perceived by researchers from different disciplines [10], diameters of the exhaled droplets can cover a broad range, from 0.1 to 1000 μm [11, 12], for which the fluid mechanics and evaporation kinetics vary substantially. Literature suggests that sneezing may generate droplets between 0.5 – 12 μm in diameter whereas breathing and speaking may result in droplets with at an averaged diameter of 1 μm [13-16].

It is therefore crucial to further understand the characteristics of aerosol droplets once they are deposited on common surfaces. Evaporation kinetics of micro and picolitre droplets of pure liquid, e.g. water or organic solvents, have been extensively studied in the past [17, 18]. However, the complex compositional nature of saliva, consisting salts, proteins, and lipids, could influence not only the evaporation process of the microdroplet, but also the location of the viruses in the final nuclei, and consequently their viability and transmissibility on solid substrates [19, 20]. Previous studies in which model respiratory liquids were used reported that the generated microdroplets would crystallise on the substrate as the result of their evaporation process [21, 22]. It is unclear whether the surface crystallisation, as part of the drying process, could inactivate the virus because they are enveloped in the crystals, or the crystals actually preserve the viruses and prolongs their viability upon rehydration. Furthermore, characteristics of the microdroplet nuclei could have a direct impact on the effectiveness of sample collection protocols in that the dissolution kinetics of the crystalline phase is much slower than the amorphous phase [23]. Finally, adhesion of microdroplet nuclei to the underlying substrate could be critical to the contact transmission.

In the present work, we investigated systematically the morphological and nanomechanical properties of microdroplet nuclei and their formation kinetics as a function of the surface in contact. Using an advanced inkjet printing setup, we were able to generate picolitre droplets and subsequently deposit them onto five inanimate common surfaces. We found that surface characteristics, including both surface energy and roughness, could influence the evaporation process of microdroplets, the structure of the resulting nuclei, and consequently the viability and transmissibility of virus.

## Methodology

### Materials

Phosphate buffered saline (PBS), mucin extracted from porcine stomach (M1778), magnesium chloride, calcium chloride, ammonium chloride and Dulbecco's modified Eagle's medium (DMEM) were purchased from Sigma-Aldrich (Somerset, UK).

### Artificial saliva

Artificial saliva, an aqueous mixture of salts, nutrients, and mucin, was prepared to replicate human saliva [24]. The addition of inorganic ions in the mixture offers osmotic balance and buffering especially during the injection process [24-26]. The artificial saliva composition used in the present work is shown in Table 1 [16, 27-29].

Table 1: Composition of artificial saliva per 1 L of aqueous solution in deionized water.

| Compound | Quantity | Compound | Quantity |
| --- | --- | --- | --- |
| $MgCl_2 \cdot 7H_2O$ | 0.04 g | $Na_2HPO4$ | 0.42 g |
| $CaCl_2 \cdot H_2O$ | 0.13 g | $(NH_2)_2CO$ | 0.12 g |
| $NH_4Cl$ | 0.11 g | Mucin | 1.00 g |
| KCl | 1.04 g | $KH_2PO_4$ | 0.21 g |
| DMEM | 1 mL | NaCl | 0.88 g |

Picolitre droplet generation

A customised Jetxpert Print Station (Imagexpert, NH) was used to generate arrays of picolitre droplets of the artificial saliva that were deposited on to various solid substrates. The print station uses a conveyor belt instead of a single linear stage to improve processing efficiency. The print head (GH2220, Ricoh, UK) was thoroughly cleaned with distilled water, followed by the prepared artificial saliva, prior to the droplet generation. To prevent any potential clogging of the print head by the drying saliva solution, an automated printing cycle that ejects 200 droplets every 5 seconds was programmed to keep the nozzles wet and prevent the accumulation of mucin.

To generate the desired droplet size, the print head driving signals were optimised in terms of voltage amplitude and pulse timing. The Jetxpert print station was equipped with a strobing camera and the relevant software to measure the droplet size. Once droplets with diameter of approximately 10 μm were produced consistently, the print head was placed over the conveyor stage to generate arrays of droplets that were approximately 175 μm apart by printing with a resolution of 300 × 300 dpi. A standard USB camera (Hayear 1136) was mounted above the printing stage to record videos (30 fps) of the printed artificial saliva droplets and to monitor the averaged drying time on each substrate. Five solid substrates of distinctively characteristics were used to represent the common surfaces: borosilicate glass slides, polytetrafluoroethylene (PTFE), stainless steel (SS), acrylonitrile butadiene styrene (ABS) and melamine.

Fluorescence Microscopy

Fluorescent microscopy was carried out to identify the residues of the deposited microdroplets. Alexa Fluor 488 Maleimide of excitation/emission spectra 493/516 nm was used for fluorescent imaging purposes. For one dedicated set of experiments, the fluorophore was dissolved in Dimethyl sulfoxide (DMSO) at a concentration of 1 mM, of which 10 μL were introduced to 10 mL of artificial saliva solution prior to the printing process. Sample images were captured after 5 days of drying, using an inverted fluorescence microscope (Olympus IX71).

Contact angle goniometer

An optical tensiometer (Theta Flow, Biolin Scientific, UK) was used to measure both advancing and receding contact angles of artificial saliva on the solid substrates, for which droplet of 4 µL was produced by the associated micro-syringe. To quantify the surface free energy (SFE) of these substrates, 2 µL diiodomethane (apolar) or water (polar) were placed on the substrates, and equilibrium contact angles were recorded subsequently.

Owens/Wendt theory describes the surface energy of a solid as having two components, a dispersive component and a non-dispersive (or polar) component. Mathematically, the theory is based on the combination of two fundamental equations (Good's equation and Young's equation) that describe interactions between solid surfaces and liquids [30, 31]. This equation has a linear form y = mx + b, wherein:

$$y = \frac{\sigma_L \cdot (cos\theta + 1)}{2 \cdot \sigma_L^{D\,1/2}}; x = \frac{\sigma_L^{P\,1/2}}{\sigma_L^{D\,1/2}}; m = \sigma_S^{P\,1/2}; b = \sigma_S^{D\,1/2} \qquad \text{(Eqn. 1)}$$

Diiodomethane has a relatively high overall surface tension of 50.8 mN m$^{-1}$ but no polar component, so that $\sigma_L = \sigma_L^D$ = 50.8 mN/m, whilst water has a surface tension $\sigma_L^P$ = 46.4 mN m$^{-1}$ for the polar component, and $\sigma_L^D$ = 26.4 mN m$^{-1}$ for the disperse component. The slope m of that line is used to calculate the polar component of the surface energy of the solid $\sigma_S^P$ and the intercept b is used to calculate the dispersive component of the surface energy of the solid $\sigma_S^D$, the overall surface free energy (SFE) of the solid being defined as SFE = $\sigma_S^D + \sigma_S^P$.

Atomic force microscopy

Surface topography and characteristics of the dried droplets were examined using an atomic force microscope (Dimension 3100, Bruker). All imaging and force curve requisition was performed with silicon nitride cantilevers (PNP-TR-Au, spring constant 0.08 N m$^{-1}$, Apex Probes Ltd. UK). All cantilevers were subject to the same cleaning routine (rinsing with ethanol followed by exposure to UV-Ozone for 15 mins) before the first use and between every two samples. Force measurements on the dried artificial saliva nuclei were acquired with a maximum loading force of 5 nN. At least 100 repetitions, in the form of a 7 × 7 matrix (200 nm apart in both x and y directions), were acquired at 2-3 random locations within the area of interest. Young's moduli of the dried products were estimated by fitting the approaching component of the force curves with Sneddon model (conical indenters):

$$F = \frac{E}{1-\nu^2} \cdot \frac{2 \tan\alpha}{\pi} \cdot \delta^2 \qquad \text{(Eqn. 2)}$$

where $E$ and $\nu$ are the Young's modulus and the Poisson's ratio (0.5 for organic compound) of the materials being indented, respectively; $\alpha$ is the semi-opening angle of the AFM tip which is

35° in the present study; $\delta$ is the indentation depth which is calculated by subtracting the cantilever deflection from the measured sample height.

## Results and Discussion

### Microdroplet generation

The Ricoh GH2220 print head deployed in the present work was chosen due to its highly adaptable performance with different types of fluid. Although the viscosity of the artificial saliva was well below the ideal operating point for the print head, there is a wide latitude to modify the print head driving signals. To measure and adjust droplet size, the camera of the Jetxpert drop watcher is triggered in sync with the GH2220 drive electronics (Meteor Inkjet Ltd, UK). This is a standard method to capture the flight trajectory of individual droplets as they are ejected from the inkjet head. The goal was to find the optimal pulse shape and voltage amplitude to actuate the piezoelectric elements of the print head, so as to eliminate satellite droplets present and reach the desired droplet size. Figure 1 presents a series of images captured, showing the trajectory of a single artificial saliva droplet of approximately 5 pL volume being ejected from the print head. We were able to capture the entire process, from moment when the droplet was detached from the nozzle (far left) and the subsequent movement in the air without any satellite drops.

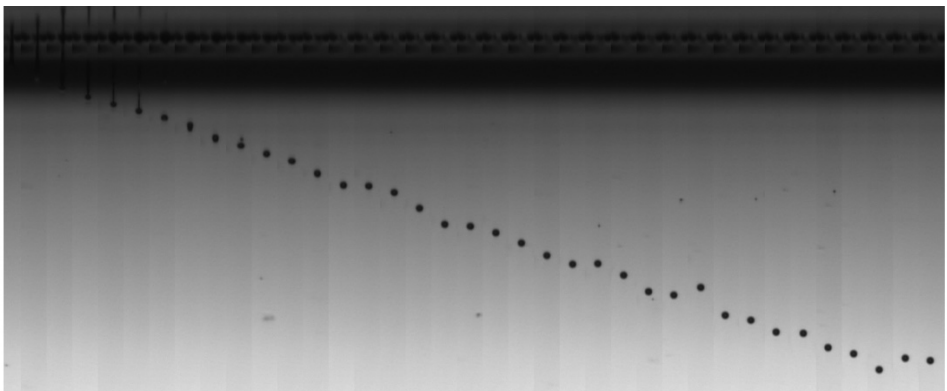

Figure 1. Mosaic of images captured from 37 separate ejection events by the jetxpert camera to show the flight trajectory of a single artificial saliva droplet from the moment of ejection (far left). Diameter of the droplet is 10 µm.

The two rows of nozzles of the GH2220 enabled printing in a single pass with a pixel addressability of 300 dpi, which could produce droplets as close together as 85 µm. To minimise the possibility of any uncontrolled merging of adjacent drops across the range of surfaces studied, a print image using a ordered dither of approximately 23% coverage was implemented, resulting an average droplet separation of ~175 µm. With these settings, and belt speed of just 5m min$^{-1}$ the print frequency was kept low (<1 kHz), thus minimising the chance of nozzle failure due to the sub-optimal rheology. A camera was positioned next to the solid substrate to capture the drying of the deposited droplets from the point the conveyor came to rest, approximately 1 second after printing.

The ability to consistently and rigorously generate microdroplets of uniform size provides a substantial advantage over the other techniques in generating aerosol droplets such as atomiser and nebuliser. It offers the assurance to compare droplet nuclei formed on different substrates, with minimal variation in the droplet characteristics.

Evaporation of the deposited droplet

Droplets of 10 µm diameter (volume is approximately 4.8 picolitre) were deposited on five different surfaces, namely: Glass, PTFE, SS, ABS and Melamine. The time it took for the droplet to evaporate on the substrates, upon deposition, was estimated using the videos captured, and is presented (Figure 2) as a function of the corresponding surface free energy (SFE) established based on de-ionised water and diiodomethane. The drying time estimated appears to correlate with the SFE of the substrates: the greater the SFE is, the longer it takes for the droplet to evaporate.

Evaporation kinetics, including time, of microlitre liquid droplets on solid substrate have been studied extensively at millimetre scale [17, 18]. It is widely accepted that there are two limiting drying scenarios: i) the contact line of the liquid/solid interface remains constant throughout the drying whilst contact angle decreases, called constant contact radius (CCR) mode; and ii) droplet radius reduces whilst the contact angle remains the same, called constant contact angle (CCA) mode, which is often observed on hydrophobic substrates [32]. It is however worth noting that variations in both the physical and chemical characteristics of the substrate could help to enhance contact line pinning, although strongly hydrophobic substrates could be exceptions [33].

For droplets of picolitre volume, Talbot and co-workers investigated the evaporation kinetics of water and ethanol droplets on surfaces with different wettability and thermal conductivity [34]. The droplet volume studied ranges from 4 to 65 pL, and the drying time for water droplets was around 4 seconds for the surfaces used. This is in a very similar time frame to that observed in the present work, which is not surprising because the artificial saliva used in our work consists of 97% water. It was not possible to measure the contact angle of the picolitre droplet in the present investigation, but previous studies suggest that it would be several degree less than that measured using microlitre droplet [34]. Although drying and evaporation kinetics of picolitre droplets are beyond the scope of this work, the observed correlation between drying time and surface free energy does not seem to agree with the previous finding that evaporation on hydrophilic substrates is faster than on hydrophobic substrates. This inconsistency could be attributed to the substrates used, or the surface adsorption of mucin molecules upon deposition. We would like to highlight that the variations in the actual drying time of artificial saliva microdroplets of 10 µm are within two seconds, which indicates that transfer of aerosol droplet in liquid state is unlikely, and suggests that the effect of surface hydrophobicity on the drying time of microdroplets (< 10 µm) is not a significant factor in the context of surface transmission. The short time (a few seconds) for these biological picolitre droplets to reach full dryness, as opposed to the length scale (mins to tens of mins) it take for micro- to macro-sized droplets, underlines the importance of focusing on the physico-chemical and virological characteristic of the dried nuclei of such, rather than solely on fresh respiratory deposition in terms of surface transmission.

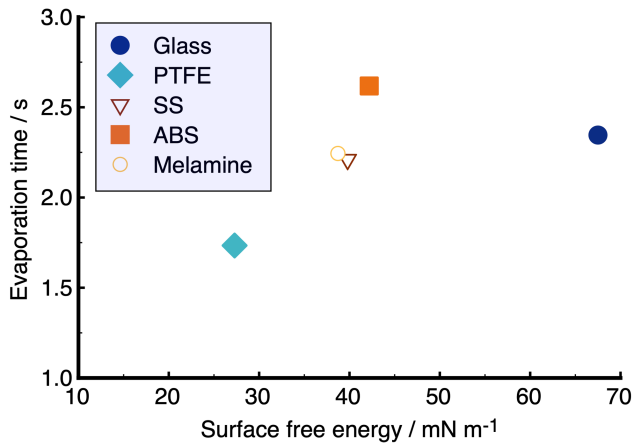

Figure 2. Evaporation time of artificial saliva microdroplet on five different substrates as a function of their corresponding surface free energy (SFE) that was calculated using diiodomethane and deionized water.

Advancing and receding contact angle measurements

Advancing and receding contact angles (CA) of artificial saliva (2 µL) were measured on glass, stainless steel (SS), PTFE, ABS and melamine, as presented in Figure 3. Of the five substrates investigated, glass shows the smallest advancing CA (42.0°) and the second smallest receding CA (15.2°) of the range, while PTFE results in the greatest ones (115.2° and 29.5° respectively), which correlates with their SFE values: 67.5 mN m$^{-1}$ and 27.3 mN m$^{-1}$ for glass and PTFE respectively. SS, ABS, and melamine present intermediate values, in terms of SFE and contact angle of artificial saliva, showing similar correlation between advancing and receding contact angle values. The advancing contact angle on SS is noticeably less than that on ABS or melamine.

The differences in SFE, advancing, and receding contact angles between the five substrates surveyed are likely due to the synergistic effect of surface characteristics, e.g. chemical nature, topography, and roughness. Glass substrate shows with the lowest roughness ($R_a$), 1.1 ± 0.1 nm (obtained by AFM images over an area of 50 × 50 µm), followed by ABS and melamine with 3.8 ± 0.7 nm and 3.7 ± 0.8 nm respectively, whilst PTFE and SS present great roughness with $R_a$ = 19.4 ± 3.3 nm and 33.6 ± 6.1 nm respectively. Although SS presents a high surface roughness, both advancing and receding CA on SS are less than those on ABS and melamine, which is likely attributed to the hydrophilic nature of SS, whilst ABS and melamine are less polar. This supports the consideration of liquid contact angle, and is consistent with our previous experimental work in which stainless steel samples of different surface finishing were investigated [35]. It is worth noting that substrates, such as SS, ABS, and melamine, whether smooth or rough, can present high advancing CA but low receding CA, suggesting that it is possible for those less polar substrates to an increased interaction with the artificial saliva once the surface is already wet, while it wouldn't have been the case when dry. At the opposite, PTFE only presents high advancing and high receding CA. The surface characteristics discussed at the macrodroplet scale, including both polarity and roughness, are still important consideration at the microdroplet scale.

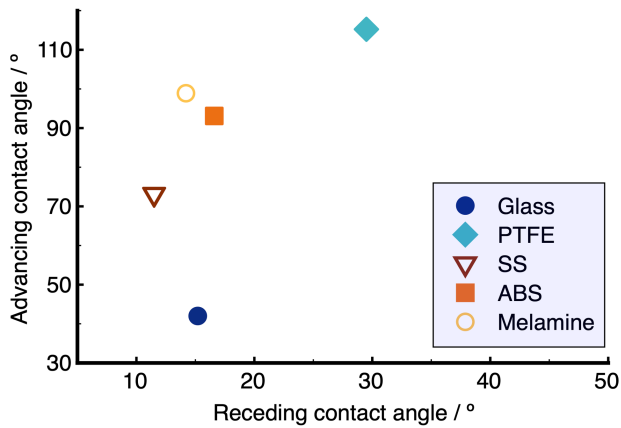

Figure 3: Advancing contact angles (CA) of artificial saliva on the five substrates as a function of the corresponding receding contact angles (CA).

Droplet nuclei characterisation

Morphology of the microdroplet nuclei with microscopic and nanoscopic spatial resolution was subsequently investigated using fluorescence microscopy and atomic force microscopy, respectively. Artificial saliva mixture containing Alexa fluor 488 Maleimide was deposited on a set of substrates using the same printing conditions and imaged by a fluorescence microscope afterwards. The acquired images are shown in Figure 4 where the array of droplet nuclei is distinctively identifiable on some of the substrates, in particularly on the glass substrate (Figure 4a & 4b). The series of solid particulates confirm the robustness of the method in preparing the microdroplet nuclei. However, they are not noticeable on substrates such as ABS (Figure 4e) and melamine (Figure 4f), which is likely due to the signal fluctuation from the background.

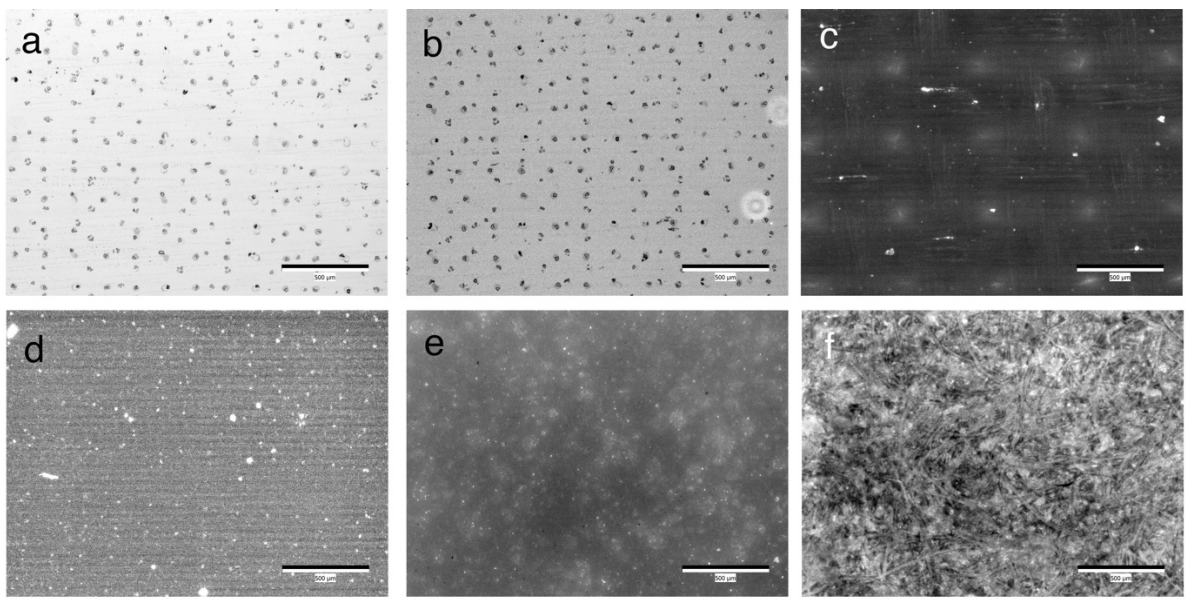

Figure 4. (a) bright field, (b) fluorescence images of artificial saliva microdroplet nuclei on glass. Fluorescence images of droplet nuclei on (c) PTFE, (d) stainless steel, (e) ABS, and (f) melamine upon deposition of artificial saliva added with fluorophore (Alexa Fluor 488 Maleimide). Scale bar corresponds to 500 μm.

To establish fine details of the microdroplet nuclei, AFM was deployed to survey different regions of the solid substrates, of which representative images are shown in Figure 5. Two distinctive morphological features were observed: large crystals surrounded by small solid residues (islands), as shown on glass, ABS, and melamine (Figures 5b, 5h, and 5j); or large crystals without any noticeable neighbouring residue (Figures 5d & 5f). As a result of the evaporation process of artificial saliva, such notable variation could be solely attributed to the characteristics of the solid substrate.

As explained in a recent study by Lieber and colleagues [36], respiratory fluid such as saliva consists of a range of inorganic salts and proteins, with a major fraction of water that will evaporate under typical ambient conditions. They used an acoustic levitator to study the temporal evolution of saliva droplets that underwent evaporation in the air, and reported that the ratio between the equilibrium and initial diameter of a droplet is 20%, assuming an initial combined mass concentration of salts and proteins of 0.8%. Although this ratio is unlikely to be applicable for the droplets deposited on a substrate, their study highlights the significant difference between the evaporation characteristics of water and saliva, and suggests that both precipitation and crystallisation could take place during the evaporation of saliva droplets.

Concerning the drying process of a surface deposited macroscopic droplet, a capillary flow induced by the evaporation of water molecules at the air/liquid interface generates a convective mass transfer phenomenon and liquid at the ridge is replenished by liquid from the interior [37]. The convective flow could carry the dispersed solutes to the solid/liquid contact line, resulting in a circle of solid deposit, known as the coffee ring effect [37]. This flow would be countered by Marangoni flows that redistribute the particles back to the centre of the droplet [38]. In the present work, no coffee ring effect was observed on any of the substrates, suggesting that either the evaporation time was not sufficient to drive the solutes, such as proteins, to the edges of the microdroplet, or the Marangoni flows were strong enough to counter the convective mass transfer. Instead, solid particles with crystallisation features were seen on all substrates surveyed, which is consistent with the results reported by Vejerano and Marr [21] who studied the transformation of a mixture of mucin, salt, and surfactants, 1,2-dihex- adecanoyl-sn-glycero-3-phosphocholine (DPPC), as a function of relative humidity. Using fluorescence labelled mucin and DPPC, they were able to demonstrate that the droplet exhibited an initial core-shell structure, with a great concentration of mucin at the shell (air/liquid interface). Considering the fast evaporation kinetics observed in the present work, it is very probable that mucin molecules were kept at the shell whilst the inorganic salts underwent the crystallisation process in the centre as the water molecules evaporated.

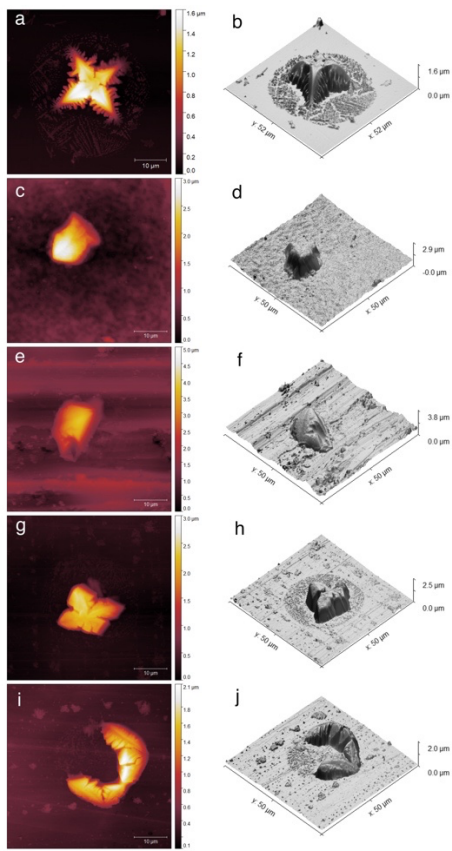

Figure 5. Morphology and the corresponding three-dimensional reconstructed images of artificial saliva droplet nuclei formed on (a,b) glass, (c,d) PTFE, (e,f) stainless steel, (g,h) ABS, and (I,j) melamine.

The dimension and morphology of these droplet residues exhibit explicitly a compliance with the corresponding surface free energies of the substrates on which the picolitre droplets landed and evaporated. AFM images in Figure 5 suggest that, upon the initial contact with the solid substrate, the proteinaceous microdroplets had a maximal contact area on substrates with high surface free energy, but kept a minimal contact with the ones with low surface free energy. For instance, the droplet residue on glass spans a width of 35 µm, which is the largest of the 5 substrates, and a peak height of approximately 1.5 µm, which is the smallest among the substrates. In contrast, the droplet nuclei on PTFE measure only approximately 10 µm in diameter but are 3 µm high, which is the tallest in this range. This observation echoes the results reported previously concerning the evaporation process of picolitre droplet [34] that water droplets followed a pinned contact line on glass but a moving contact line on PTFE, which results in droplet nuclei of different morphology and crystalline phase.

Glass, ABS, and melamine possess minimal surface roughness and low receding contact angles (Figure 4), on which the residue of artificial saliva microdroplets were found in the form of small dendrites or islands across the microdroplet residue region, next to the large crystal (Figure 6). This correlates with the low receding CA measured on the same substrates at the millimeter scale, which evidences that constant contact radius (CCR) mode is appropriate for the evaporation of respiratory microdroplet on substrates with high surface free energy. While the initial spreading of a microdroplet on the solid substrate is reflected by the advancing contact angle, it is the receding contact angle that plays a critical role in the subsequent evaporation process. Although SS possesses low receding contact angle (Figure 4), only a large crystal was found in the droplet region, which is probably due to either its excessive roughness in comparison with the other substrates tested or the anisotropic nature of the surface finishing. PTFE has a less surface roughness than SS, and a similar morphology (crystal only) can be seen in Figure 5c and 5e – this is because PTFE has a low surface free energy and high receding contact angle (Figure 4). We can safely conclude that constant contact angle (CCA) mode of evaporation is applicable here.

The contrast between evaporation modes of respiratory fluid on solid substrates not only demonstrates the effect of surface characteristics on drying, but could have significant implications on virus stability in the context of surface transmission and virus detection. Evaporation of water in such a fast timeframe could drastically change the phase and local concentration of virus, protein, and salt. Some previous study suggest that mucin could act as a protective matrix to help virus surviving several days in a completely dried condition [39]. Similar findings were reported for both SARS-CoV and MERS-CoV that could be recovered from inanimate substrates after days, weeks, or months [40, 41]. The effect of fast transformation with the microenvironment of each respiratory droplet on the structural integrity of a virus is beyond the scope of this work. However, our work shows that appropriate antiviral surface strategies could be developed to inactivate or disrupt virus encapsulated in microdroplet nuclei by modulating the evaporation process on surface. For example, surface deposited surfactants could be used to influence the Marangoni flow and consequently the evaporation mode. It is equally possible to adjust surface free energy of a substrate once we find out whether crystalline or amorphous phase is more effective in disrupting the virus envelope.

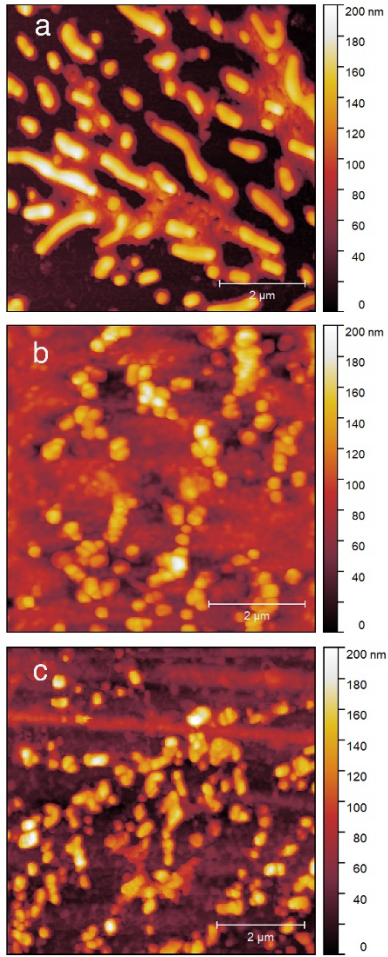

Figure 6. Small islands of nuclei formed aside the large crystal on (a) glass, (b) ABS and (c) melamine. Scale bar is 2 μm.

### Nanomechanical properties of the droplet nuclei

Nanomechanical measurements were performed to further evaluate the proposed 'core-shell' structure, during which a nanoscopic tip (radius in the region of 10 nm) made contact with the microdroplet nuclei and retract subsequently at a given frequency (1 Hz in the present work). Since the experiments were carried out in ambient, capillary force between the AFM tip (silicon nitride) and the nuclei dominates the surface adhesion (recorded as the hysteresis between approaching and retraction curves in Figure 7a), which can be described as

$$F_{cap} = 4\pi R \gamma_L \cos\theta \qquad (Eqn. 3)$$

where $R$ is the AFM tip radius, $\gamma_L$ is the surface tension of water, and $\theta$ is the contact angle of water on the two surfaces in contact. It is clear that the capillary force is determined by the local surface free energy of the nuclei that engages with the AFM tip. Used in the past to evaluate the

SFE of mineral such as calcite with nanoscopic spatial resolution [42], this method was utilised in the current study to survey various locations across the microdroplet residue regions, including both the large crystals and the small islands, found on all five inanimate substrate. Equally, AFM based force spectroscopy could be used to identify any surface heterogeneity since polar chemical moieties would attract water molecules, which consequently result in an increased surface adhesion [43]. At least 100 force curves, with a representative one shown in Figure 7a, were collected from everyone point examined to ensure statistical robustness.

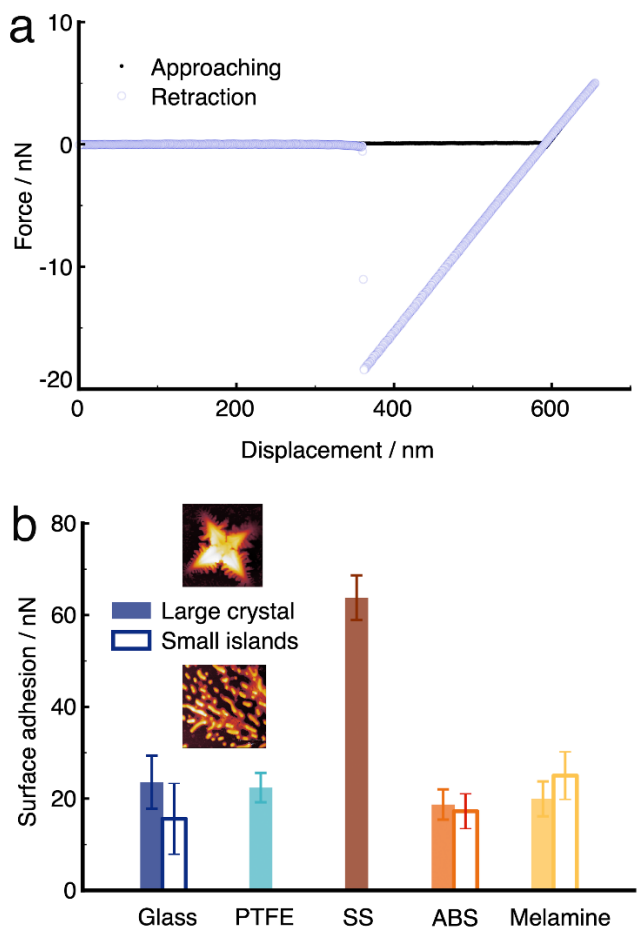

Figure 7. Adhesion forces between the AFM tip and different surfaces. Measurements were taken on both the large crystals and the small nuclei area. No small islands were observed on PTFE and SS surfaces.

Averaged surface adhesion values are presented in Figure 7b. It shows that very similar magnitudes of surface adhesion were acquired from the microdroplet nuclei deposited on most of the substrates, except SS. The consistent values of surface adhesion on glass, PTFE, ABS, and melamine, confirm that the surface free energy of droplet nuclei is very similar, supporting that our proposed 'core-shell' model is applicable on these surfaces. The absolute values (around 20 nN) reported agrees with a previous study whereby surface adhesion between a silicon AFM cantilever and a glass slide was measured [44] as well a recent work by ourselves concerning surface adhesion on hair fibres in ambient conditions [45]. The nuclei present on SS resulted in a surface adhesion that is approximately three times more than those on the other inanimate substrates, suggesting that they possess high surface free energy. We speculate that this might be attributed to the exposure of hydrophilic region of mucin on the surface as the result of evaporation of the proteinaceous microdroplets.

By indenting the AFM cantilever into the entities present on the solid substrate (droplet nuclei in the present work) for tens of nm, it is possible to quantify the viscoelasticity of the nuclei, and to evaluate their structural characteristics. Young's moduli of the nuclei, both large crystal and small islands regions, are presented in Figure 8. The locations surveyed, across all five substrates, show values of close range (~ 4 MPa), except the small islands on ABS. The Young's modulus values are consistent to those acquired on polymer film [46], which further supports the proposed structure that mucin molecules present as the shell for the microdroplet nuclei. The high Young's modulus acquired from the small islands on ABS could be attributed to the variation in the concentration.

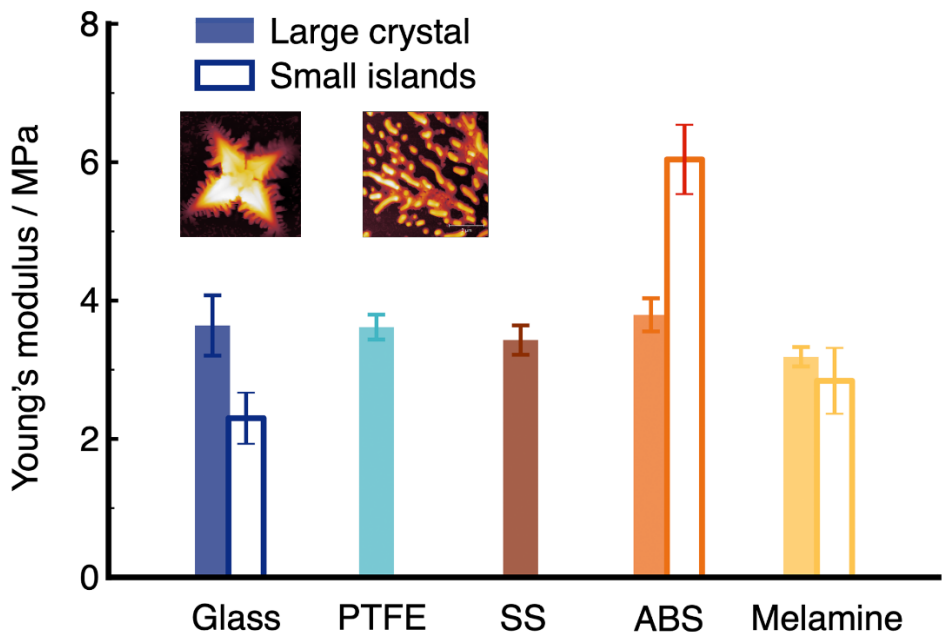

Figure 8. Young's modulus as function of surface materials receiving droplets. Both large crystals (filled columns) and small nuclei (empty columns) were surveyed.

## Summary

An advanced inkjet printing method was used to generate microdroplets of respiratory fluid on five common substrates. We show that surface characteristics of the inanimate substrate have substantial impact on not only the drying kinetics of respiratory fluid microdroplets, but also the properties of the resulting nuclei. The evaporation kinetic of artificial saliva follows constant contact radius mode on substrates with low surface free energy or great roughness, but constant contact angle mode on those with high surface free energy. This results in two distinctively different morphology of the microdroplet nuclei. Atomic force microscopy-based methods, force spectroscopy and nanoindentation, were deployed to investigate the nuclei on all five substrates, which could be an invaluable approach in the future studies of respiratory droplet. The nanomechanical measurement results support a core-shell structure of the microdroplet nuclei due to the fast evaporation process of microdroplet, which could have significant implication on the surface viability/transmissibility of virus as the crystals might preserve the integrity of viruses, or disrupt the virus structure. The work highlights the importance of length scale on the drying of droplets and consequently the possibility of surface transmission of virus containing small droplets.

## Acknowledgement

We would like to thank Dr Huaiyu Yang, Department of Chemical Engineering, Loughborough University, for discussion related to surface crystallisation phenomena. We would like to thank Drs Richard Thomas, Maurice Walker, and Jack Vincent, Defence Science and Technology Laboratory, for fruitful discussion.

## Funding

This work was supported by the Engineering and Physical Science Research Council [grant number EP/V029762/1]. ZJZ thanks the Royal Academy of Engineering for an Industrial Fellowship award (IF2021\100).